\documentclass[12pt]{article}
\usepackage{epsfig}
\begin{document}

\begin {center}
{\Large \bf A fresh look at $\eta_2(1645)$, $\eta_2(1870)$,
$\eta_2(2030)$ and $f_2(1910)$ in $\bar pp \to \eta \pi ^0\pi ^0\pi ^0$}
\vskip 5mm
{A V Anisovich $^ c$, C J Batty $^b$, D V Bugg $^a$, V A Nikonov
$^c$ and A V Sarantsev $^c$.\\
{\normalsize $^a$ \it Queen Mary, University of London, London E1\,4NS,
UK} \\
{\normalsize $^b$ \it Rutherford Appleton Laboratory, Chilton, Didcot,
OX11 OQX, UK} \\
{\normalsize\it $^c$ St. Petersburg Nuclear Physics
Institute, Gatchina, St. Petersburg district 188350, Russia} \\
[3mm]}
\end {center}
\date{\today}

\begin{abstract}
There is a large discrepancy between results of Crystal Barrel and
WA102 for the branching ratio $R =
BR[\eta_2(1870) \to a_2(1320)\pi]/
BR[\eta_2(1870) \to f_2(1270)\eta]$.
An extensive re-analysis of the Crystal Barrel data redetermines
branching ratios for decays of $\eta _2(1870)$, $\eta _2(1645)$,
$\eta _2(2030)$ and $f_2(1910)$.
This re-analysis confirms a small value for $R$ of $1.60 \pm 0.39$,
inconsistent with the value $20.4 \pm 6.6$ of WA102.
The likely origin of the discrepancy is that the WA102 data contain
a strong $f_2(1910) \to a_2\pi$ signal as well as $\eta _2(1870)$.
There is strong evidence that the $\eta _2(1870)$ has resonant
phase variation.
A peak in $f_2(1270)a_0(980)$ confirms closely the parameters of the
$a_2(2255)$ resonance observed previously.
A peak in $\eta_2(2030)\pi$ is interpreted naturally in terms of
$\pi_2(2245)$ with reduced errors for mass and width
$M=2285 \pm 20(stat) \pm 25 (syst)$ MeV, $\Gamma = 250
\pm 20 (stat) \pm 25 (syst)$ MeV.

\end{abstract}

\section {Introduction}
The objective of this paper is to re-examine data from Crystal Barrel
(CBAR) \cite {Cooper} \cite {Adomeit} \cite {eta3pi}
concerning $\eta _2(1645)$, $\eta _2(1870)$, $\eta _2(2030)$ and
$f_2(1910)$.
One motivation is to study a large discrepancy
for branching ratios of $\eta _2(1870)$.
Earlier CBAR work found a branching ratio
\begin {equation}
R = \frac {BR[\eta _2(1870) \to a_2(1320)\pi ]}
{BR[\eta _2(1870) \to f_2(1270)\eta ]} = 1.27 \pm 0.17,
\end {equation}
see Table 5 of Ref. \cite {eta3pi}.
This is much smaller than the value $20.4 \pm 6.6$ claimed by the
WA102 collaboration in central production of $\eta \pi \pi$
\cite {WA102A}.
Other branching ratios determined by the CBAR data are
redetermined here.
This re-analysis incorporates many further details of spectroscopy in
this mass range which have appeared since the year 2000.

A review of these earlier publications will set the scene and
introduce the relevant resonances.
For a comprehensive review  of the CBAR work with $\bar p$ in flight
see Ref. \cite {review}.
The first study of $\eta _2(1645)$ and $\eta _2(1870)$ in CBAR data
was at two beam momenta: 1200 and 1940 MeV/c \cite {Cooper}
\cite {Adomeit}.
The $\eta _2(1645)$ was observed decaying to $a_2(1320)\pi$
and $a_0(980)\pi$.
No other $ a_0$ or $a_2$ appear in the data, so  $a_0$
will be used hereafter as a shorthand for $a_0(980)$ and likewise $a_2$
for $a_2(1320)$.

There was also a strong $f_2(1270)\eta$ signal near
its threshold $\sim 1810$ MeV.
It could not be explained as the high mass tail of $\eta _2(1645)$.
The reason was that a single $\eta _2(1645)$ decaying to $f_2\eta$
would contain a large $f_2\eta $ signal in both numerator and
denominator of the Breit-Wigner amplitude;
cancellation between numerator and denominator cannot accomodate the
large $f_2(1270)\eta$ signal.
The data were fitted with the addition of the $\eta
_2(1870)$ though the data did not rule out the possibility of a
non-resonance threshold effect at that time.
\begin{figure}[htb]
\begin{center}
\vskip -8mm
\epsfig{file=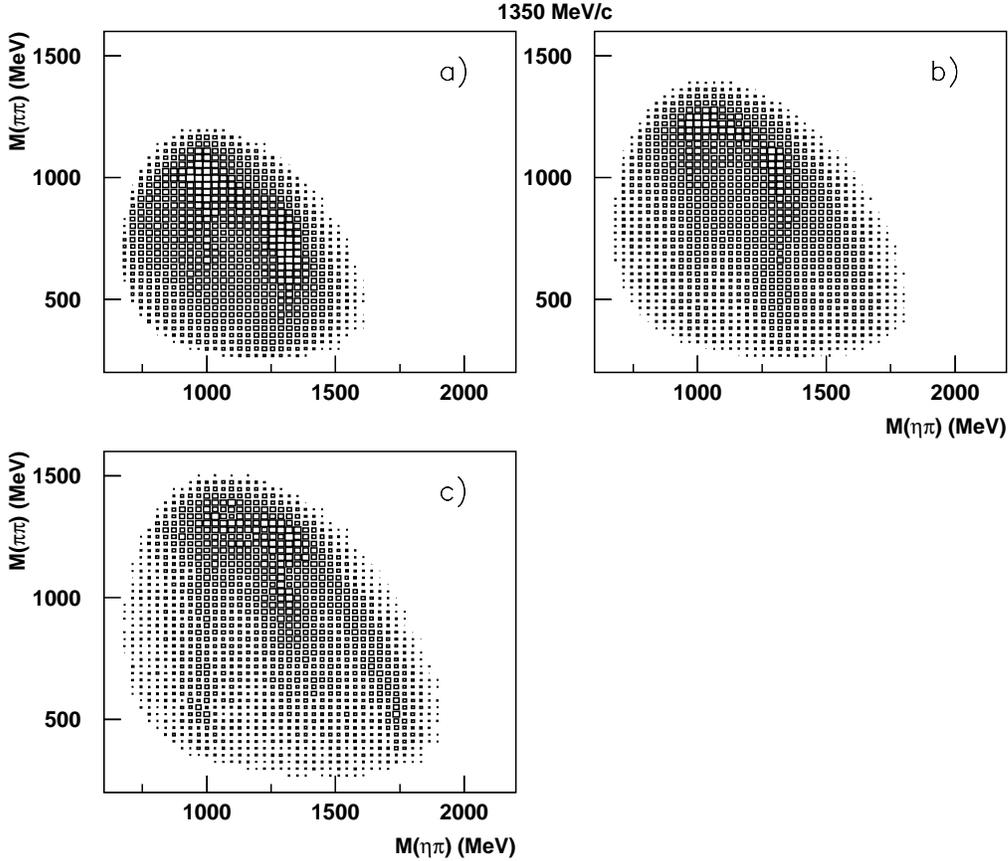,width=15cm}
\vskip -9mm
\caption {Scatter plots of $M(\pi \pi)$ v. $M(\eta \pi)$ for
three ranges of $M(\eta \pi \pi)$:
(a) 1560-1750 MeV, centred on $\eta_2(1645)$,
(b) 1775-1945 MeV over $\eta _2(1870)$ and
(c) 1945-2115 MeV, centred on $\eta _2(2030)$.
The beam momentum is 1350 MeV/c.}
\end{center}
\end{figure}

The majority of CBAR data with $\bar p$ in flight were taken in 1996.
For $\eta 3\pi ^0$, statistics were a factor 7 higher than earlier
data at each of nine beam momenta ranging from 600 to 1940 MeV/c,
i.e. an overall increase of statistics by a factor $\sim 30$.
There are typically 25--30K events at each beam momentum and a total
of 213K events.

Fig. 1 shows scatter plots from CBAR data in three ranges of
$\eta \pi \pi$ mass.
Fig. 1(a) shows the $\eta_2(1645)$ mass range; there is a vertical
band due to $a_2\pi$ and also a peak in $\pi \pi$ near 1 GeV.
The spin analysis ruled out $f_0(980)$, showing
that the peak in $\pi \pi$ is due to the low mass tail of $\eta
_2(1870) \to f_2(1270)\eta$.
Fig. 1(b) shows the $\eta _2(1870)$ mass range.
The $a_2(1320)$ and $f_2(1270)$ bands appear of similar strength;
however, the $f_2$ is somewhat broader and therefore stronger.
There is clearly no large excess of $a_2$ decays in this mass interval.
The branching ratio quoted by WA102 would require an
$\eta_2(1870) \to a_2\pi$ signal a factor $\sim 3.5$ larger than
$f_2(1270)\eta$ when one allows for relative decay rates of
$a_2\to \eta \pi$ and $f_2(1270) \to \pi ^0\pi ^0$ and for charge
combinations.
Fig. 1(c) shows the mass range of $\eta_2(2030)$; there is a peak
where $a_2(1320)$ and $f_2(1270)$ bands cross.
The $a_2\pi$ band is strong and $f_2(1270)\eta$ is weaker.
Note also that weak $a_0(980)$ bands are visible in all three panels.

There is additional evidence for $\eta _2(2030) \to f_2(1270)\eta$ and
$a_2\pi$ in further CBAR data for
$\bar pp \to \eta \pi ^0\pi ^0$, see Fig. 11 of Ref. \cite {epp}.
A distinctive feature of this state is its strong decay to
$[a_2\pi]_{L=2}$, where $L$ is the orbital angular momentum
in the decay.

Following the year 2000 publication of $\eta 3\pi ^0$ results, a
combined analysis was made of 10 sets of data with isospin
$I=0$ and $C$-parity $+1$.
Six of these were CBAR data for
$\bar pp $ in flight $\to \pi ^0 \pi ^0$, $\eta \eta$, $\eta \eta '$,
$\eta \pi ^0 \pi ^0$, $\eta '\pi ^0\pi ^0$ and $\eta \eta \eta$ \cite
{Combined}.
The other four  were high quality differential
cross sections and polarisations for $\bar pp \to \pi ^+\pi ^-$ from
two experiments: Eisenhandler et al. \cite {Eisenhandler} and PS172,
Hasan et al. \cite {PS172}.
This analysis revealed two towers of
resonances centred at $\sim 2000$ and 2270 MeV, with all $J^P$ for
$q\bar q$ states expected in this mass range.
Most  have been observed in at least three sets of data.
Polarisation data provide a clean separation of $\bar pp$ $^3P_2$
and $^3F_2$ states.
The $f_2(1910)$ of the PDG \cite {PDG} was confirmed and identified as
dominantly $^3P_2$; a neighbouring $^3F_2$ state was observed at 2001
MeV. Further $f_2(2240)$ $^3P_2$ and $f_2(2295)$ $^3F_2$ states were
also observed. In the new analysis reported here, the first three of
these $f_2$ states play a significant role. The $f_2(1910)$ lies close
to $\eta _2(1870)$ and is important for the discussion of WA102 data.

In 2001, a combined analysis was also made of data on $I=1$, $C=+1$
states \cite {I1}.
The spectrum of states is less complete than for $I=0$, $C = +1$,
because of the lack of polarisation data.
Nonetheless, an $a_2(2255)$ appeared clearly in three channels of
data.
Secondly, there is a less well identified $\pi _2(2245)$.
These two states now appear in the analysis reported here.
So, in summary, the picture has developed significantly since the
earlier analysis of $\eta 3\pi ^0$.

The $\eta 3\pi ^0$ channel may appear to be a
complicated channel to analyse, because of the multiplicity of
$\eta \pi ^0$ and $\pi ^0 \pi^0$ combinations.
However, $\eta_2(1645)$ and $\eta_2(1870)$ were found here; also, once
one knows the mass and width of $\eta_2(2030)$ from $\eta \pi \pi$
final states, it is easily detected in the present data via its
strong decay to $[a_2\pi]_{L=2}$, which has a very distinctive angular
dependence.
Interferences between channels provide intricate
information on identifiable resonances, even though the angular
correlations cannot be displayed because they are multi-dimensional.
It is necessary to work from log likelihood and mass projections of
$\eta \pi$, $\pi \pi$, $3\pi$ and $\eta \pi \pi$.
A valuable check on the analysis is to introduce deliberate errors
into angular dependence of amplitudes; genuine signals then drop to low
values.

Two alternative starting points have been adopted.
The first is the year 2000 analysis, which now improves.
The second starts from the WA102 ratio for $\eta _2(1870) \to a_2\pi$
and $f_2(1270)\eta$.
This gives a considerably worse fit.
After iterations, the two fits converge to a single solution except for
minor points of ambiguity in $a_0\pi$ decays.
No significantly different solutions have been found at any beam
momentum despite searches from a variety of initial parameters.

The layout of the paper is as follows.
Sections 2 and 3 go through technical details of the analysis
procedure.
It may be useful to glance first at figures of later sections, so as
to appreciate the rationale for the steps discussed in Sections 2
and 3; the techniques need to be adapted to what is found in the data.
Section 2 introduces the channels which are required and deals
with formulae.
These are well known from earlier literature, but need
to be documented.
The one point requiring special treatment is the opening of the
$f_2(1270)\eta$ threshold, close to $\eta _2(1870)$.
It is necessary to fold the width of $f_2(1270)$ into the phase
space for the $f_2\eta$ final state appearing in the
Breit-Wigner amplitude.
Secondly, the way the spin dependence is treated in terms of partial
waves is discussed.

Section 3 presents features of the data.
Fig. 4 shows mass spectra for $3\pi$, $\pi \pi \eta$, $\pi \eta$
and $\pi\pi$ at one representative momentum, 1642 MeV/c; other momenta
show similar features and Fig. 1 of Ref. \cite {eta3pi} presents
spectra at 1800 MeV/c.
This section gives more detailed comparisons with data in further
figures.

Section 4 then presents essential results.
Table 1 shows changes of log likelihood when each channel is
dropped from the fit and all others are re-optimised.
This Table identifies directly the important channels and their
significance levels, and how they vary with beam momentum.
Fig. 7 presents the cross sections for all
reactions as a function of beam momentum; these lead to considerable
insight into the physics.
Peaks shown in Fig. 8 may be identified with $a_2(2255)$ and $\pi
_2(2245)$.

A revised set of branching ratios is derived for $\eta_2(1645)$,
$\eta _2(1870)$, $\eta _2(2030)$ and $f_2(1910)$.
The strong decay modes change little, but
there are some significant changes from the earlier publication
in weak channels, for reasons which are understood.
The important ratio for $\eta _2(1870) \to a_2(1320)\pi$ and
$f_2(1270)\eta $ changes only slightly and remains completely
inconsistent with the WA102 result.

Section 5 therefore re-examines WA102 mass projections for
$\eta _2(1645)$ and $\eta _2(1870) \to a_0\pi$, $a_2\pi$ and
$f_2(1270)\eta$.
There is good agreement with CBAR data for $\eta _2(1645) \to
a_2\pi$ and $a_0\pi$ and  their ratio of intensities.
There is also reasonable agreement for the line-shape of
$\eta _2(1870) \to f_2(1270)\eta$.
So masses and widths of these states agree well between the two
experiments.
The evidence for the controversial $f_2(1870) \to a_2\pi$ signal
rests on a small bump in the $a_2\pi$ mass spectrum in WA102 data.
It now appears likely that some or all of this bump is instead due to
$f_2(1910)$.
This state has a strong decay to $a_2\pi$ and weaker
decay to $f_2(1270)\eta$.
An earlier WA102 publication in fact claimed to observe $J^P=2^+$ peaks
near 1900 MeV in central production of $a_2(1320)\pi \to \rho \pi\pi$
and in $f_2(1270)\pi\pi$ \cite {WA102B}.
Section 6 presents evidence that $\eta_2(1870)$ has resonant phase
variation.
Section 7 summarises results and draws conclusions.

\section {Methodology and formulae for fitting data}
The data are fitted by the maximum likelihood method, i.e. fitting
every individual event without binning.
Log likelihood is normalised so that a change of 0.5 corresponds
to a change in $\chi^2$ of 1.
For the high statistics available here, log likelihood
follows the $\chi^2$ distribution closely as the number of
variables is varied.
The following channels are fitted:
\begin {eqnarray}
\bar pp &\to& f_2(1270)a_0(980) \\
         & \to &  a_2(1320)\sigma \\
         & \to &  \pi _2(1670)\eta  \\
         & \to &  f_1 (1285)\pi  \\
         & \to &  \eta (1440)\pi   \\
         & \to &  \eta_2(1645)\pi \\
         & \to &  \eta_2(1870)\pi  \\
         & \to &  \eta_2(2030)\pi  \\
         & \to &  f_2(1910)\pi  \\
         & \to &  f_2(2001)\pi  \\
         & \to &  f_2(2240)\pi  .
\end {eqnarray}
Here $\sigma$ stands for the $\pi \pi$ S-wave amplitude.
An incoherent phase space background is also included.
It arises from experimental cross-talk between the $\eta 3\pi ^0$ final
state and other final states, e.g. $4\pi ^0$ and
$\eta \eta \pi ^0 \pi ^0$.
This background is known accurately and is discussed in the
earlier publication \cite {eta3pi}; it is in the range 7.7-9.6\%,
increasing slowly with  beam momentum.

We can dispose of channels (4) and (5) quickly.
Their $\pi \pi \eta$ peaks are narrow and have no significant impact
on other channels.
The $f_1(1285)$ is fitted as decaying purely to $a_0\pi$.
Its mass and width need tuning by a few MeV to fit the height
and width of the observed  peak.
The $\eta (1440)$ is fitted with decays to $a_0\pi$ and $\eta \sigma$,
with interference between them.
There is also evidence for $\eta (1440) \to f_0(980)\eta$, discussed
in Ref. \cite {f0eta} and confirmed in \cite {Bes2}.
The $\eta (1440)$ appears clearly only at low beam momenta up to
1200 MeV/c and the fit to it does not change significantly from
that  reported earlier.

Tests have been made for additional resonances produced in
$\bar pp \to X + \pi$  where $X$ has quantum numbers $J^P = 0^-$,
$1^+$, $3^+$ or $4^+$.
There is no significant evidence for any of these.
The high spin states would be easily detectable from their
strong angular dependence.
Adding $0^- \to a_0\pi$ and $f_0(980)\eta$ does give a small
improvement in log likelihood, typically 20, but this is because these
amplitudes have no angular dependence and are prone to
picking up noise; there is no discernable optimum
in log likelihood as $\pi \pi \eta$ mass is varied.
The final fit omits $0^-$ states other than $\eta (1440)$.

\subsection {Treatment of partial waves}
Most channels involve a two-stage process $\bar pp \to X +\pi$,
$X \to Y + \pi$.
The orbital angular momentum in the decay to $X + \pi$ will be denoted
by $\ell$ and that in the subsequent decay to $Y + \pi$ by $L$.
The $\bar pp$ initial state is a mixture of spin singlet and triplet
partial waves.
The total spin $S$ is limited to $S_z = \pm 1$ or 0 along the beam
direction; $S_z = \pm 1$ give identical angular distributions.
A problem in the analysis is the absence of polarisation
data.
The consequence is that triplet contributions to $X$ are not cleanly
separated between $J = \ell$, $\ell \pm 1$ and $\ell \pm 2$.
It is therefore not possible to do a full partial
wave analysis of both production and decay.
This would become possible if data were available at a future date
from a polarised target.

\begin{figure}[htb]
\begin{center}
\vskip -12mm
\epsfig{file=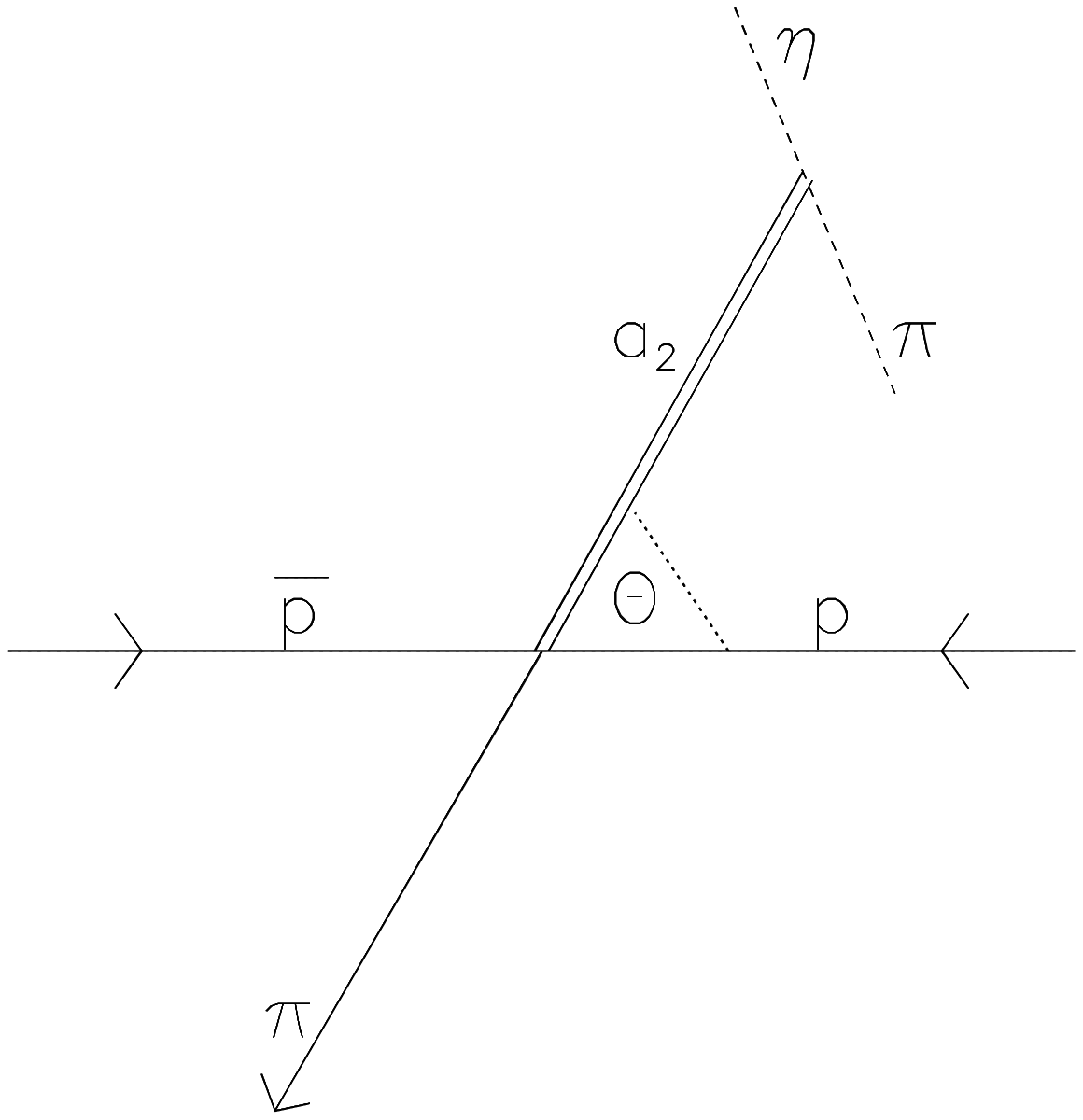,width=8cm}
\vskip -20mm
\caption {A sketch of $\bar pp \to a_2(1320) \to \eta \pi$.}
\end{center}
\end{figure}
Fig. 2 sketches the process of production and decay.
The $a_2$ and recoil pion are drawn in the $\bar pp$ rest frame.
The decay of the $a_2$ is shown after a Lorentz transformation
to the $a_2$ rest frame.
One way of writing amplitudes is to use rotation matrices to express
the initial state $|J, J_z>$ in terms of a linear combination quantised
along the $a_2$ direction.
This combination is invariant under a Lorentz boost to the rest frame
of the $a_2$ \cite {Leader}.
Then the  $a_2$ decay amplitude may be expressed in terms of the usual
Legendre polynomials.
The problem with this approach is that the rotation
matrices at the first step depend on $J$.

This problem may be avoided by a procedure known as the  Wick rotation.
After a Lorentz boost to the rest frame of the $a_2$,
a rotation of axes is made through an angle $-\theta$, the same as
in the production process but with opposite sign.
Rotation matrices then cancel between the first and second steps.
The Wick rotation alters the angles of $\pi$ and $\eta$ in the $\bar
pp$ rest frame, because of the Lorentz boost.
It preserves the fact that the initial state is restricted to
$J_z = 0$ or 1.
Since $J$ is not known, it is however necessary to discard the
angular dependence of the production process or parametrise it
empirically.
The way the programme is written, the Wick rotation is made for
every channel and every event just once, and the required
amplitudes are stored.
This speeds up the analysis by a large factor.

Two improvements of details over the earlier analysis are introduced.
For singlet states, $S_z = 0$.
The processes $\bar pp \to X + \pi$ may go via emission of a pion with
orbital angular momentum $\ell  \ge 1$, because of the pseudoscalar
nature of the pion.
For low momenta of the spectator pion, several channels rise steeply,
see Figs. 7 and 8 below.
This is consistent with P-state production.
If the $\eta _2\pi$ final states in reactions (7)--(9) are
produced via P-state pion emission, the initial state is restricted to
$J^{PC} = 2^{-+}$ unless it is exotic ($J^{PC} = 3^{-+}$ or
$1^{-+}$).
It would be surprising if exotics couple to $\bar pp$ and there is no
evidence for such exotics in other CBAR data in flight.
This leads to the useful restriction that the initial state is spin
singlet, with $S_z = 0$.
A further point is that
$\bar pp \to J^{PC} = 2^{-+} \to [2^{-+} + \pi ]_{L=1}$ has
Clebsch-Gordan coefficients such that the final state is purely $|J'=2,
J'_z=\pm 1, L =1, L_z = \mp 1 >$, where $J'$ is the spin of the $\eta
_2$. This leads to a distinctive angular dependence for the whole
amplitude describing both production and decay. It is helpful in
isolating the process $\bar pp \to \pi_1(2245) \to [\eta
_2(2030)\pi]_{L = 1}$. However, in addition we detect some significant
production from initial spin triplet states, particularly $\bar
pp~J^{PC} = 2^{++} \to [2^{-+} + \pi]_{L=0}$.

Interferences between all channels are included.
However, several spin triplet states produced from $\bar pp$
may feed a single final state such as $[f_2\pi ]_{L=1}$.
As a result, interferences between channels are not fully coherent.
To accomodate this detail, each interference term is multiplied by a
coherence factor which is allowed to optimise in the range $\pm 2$.
There are also interferences between two $a_2\pi$ and two $a_0\pi$
combinations for each resonance in $\eta \pi \pi$.
These interferences are fully coherent for a single resonance.

\subsection {The treatment of phase space}
The $\eta _2(1870)$ lies close the the $f_2(1270)\eta$ threshold.
The intensity of the $f_2\eta$ decay needs to be parametrised so as to
include the line-shape of the $f_2$ into the available phase space.
The formula for the general case $\bar pp \to X + Z$, where $X$ and
$Z$ both have significant width, is given by Eq. (40) in Ref.
\cite {Zou}.
This formula is used for channels (2) and (3), $\bar pp \to
f_2(1270)a_0$ and $a_2\sigma$.
For the simpler case of $\eta _2(1870) \to f_2(1270)\eta$,
it reduces to
\begin {equation}
\rho (f_2\eta, s) = \int _{4m^2_\pi}^{(\sqrt {s} - m_\eta)^2}
\frac {ds_1}{\pi}
\frac {4|p|}{\sqrt {s s_1}}
\frac {M\Gamma (s_1)}{(M^2 - s_1)^2 + (M\Gamma (s_1))^2} FF(s),
\end {equation}
where $p$ is the momentum of the $\eta$ in the $f_2\eta $ rest
frame;
$\sqrt {s}$ and $\sqrt {s_1}$ are the corresponding masses of
$f_2\eta$ and $f_2$.
Also $FF(s)$ is a form factor for $\eta _2(1870) \to f_2(1270)\eta$.
It is taken as a Gaussian $\exp (-\alpha p^2)$, where $\alpha = 4.5$
(GeV/c)$^{-2}$, corresponding to a radius of interaction 0.73 fm for
the overlap of $f_2$ and $\eta$.
From a wide range of CBAR and other data, $\alpha$ is known with an
error of $\pm 1.0$ (GeV/c)$^{-2}$.
In the range of the present data, results vary little over the
range $\alpha = 3.5 - 5.5$ (GeV/c)$^{-2}$.
However, the exponential dependence may be an approximation.

\begin{figure}[htb]
\begin{center}
\vskip -12mm
\epsfig{file=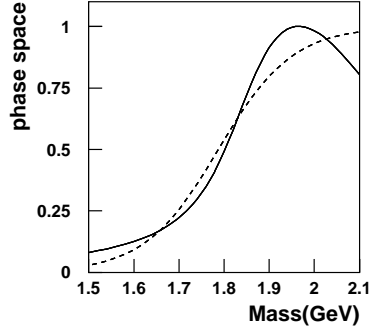,width=6cm}
\vskip -8mm
\caption {$f_2(1270)\eta$ phase space (full curve) and an approximation
with a Fermi function (dashed).}
\end{center}
\end{figure}
Fig. 3 shows  $f_2\eta$ phase space v. mass.
It peaks at $1.96 \pm 0.03$ GeV, and rises through half-height at 1.80
GeV.
It is desirable to include this $s$-dependence into the line-shape
of $\eta _2(1870)$ at least approximately.
This is done by approximating the phase space of
$f_2\eta$ by a Fermi function shown by the dashed curve of Fig. 3:
\begin {equation}
F(s)  \propto 1.0/(1.0 + 1.4\exp(5.11(1.76^2 - s))).
\end {equation}
Then the Breit-Wigner amplitude for $\eta _2(1870) \to f_2\eta$ is
\begin {equation}
f \propto \frac {\rho (f_2\eta, s)}{M^2 - s - iM[\Gamma_1 + \Gamma _2
F(s)]},
\end {equation}
where $\Gamma _1$ and $\Gamma_2$ are constants describing
decays to (1) $a_2\pi$ and $a_0\pi$, (2) $f_2\eta$.
Above 1.96 GeV, the Fermi function may be an approximation, but the
line-shape of $\eta_2(1870)$ is falling fast there.
In principle, this could lead to ambiguities in fitting the
$\eta_2(2030)$, but we find that this state is produced in a
different range of beam momenta, so in practice there is no problem.

A further potential complication is that $(M^2 - s)$ of the
Breit-Wigner denominator should strictly be replaced by
$(M^2 - s - m(s))$ with
\begin {equation}
m(s) = \frac {s - M^2}{\pi}\, P \int
\frac {M \Gamma _2 F(s') ds'}{(s' - s)(s' - M^2)};
\end {equation}
$m(s)$ is the `running mass, which makes the formula fully analytic
\cite {Sync}.
At a sharp threshold, $m(s)$ peaks strongly at the threshold.
However, for a threshold as wide as $f_2\eta$, its effect is small
and can be absorbed into optimised values of $M$ and $\Gamma_2$.

\begin{figure}[h]
\begin{center}
\vskip -10mm
\epsfig{file=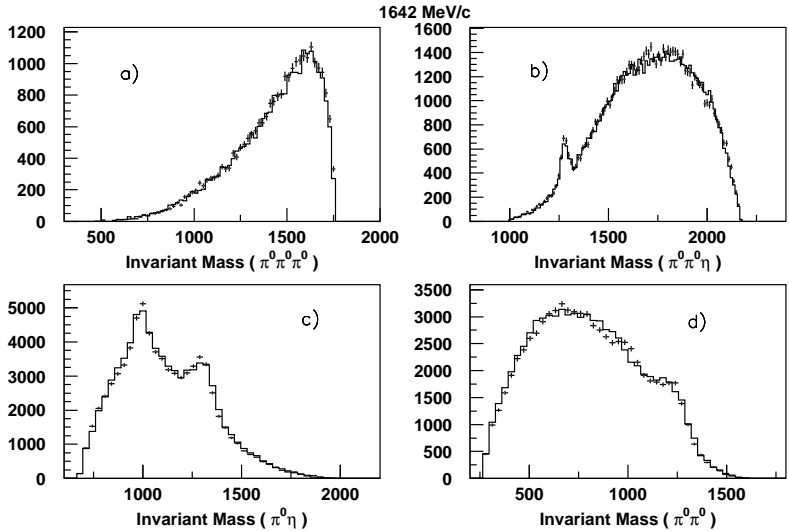,width=12cm}
\vskip -6mm
\caption {Mass spectra for (a) $3\pi$, (b) $\eta \pi \pi $,
(c) $\eta \pi $ and (d) $\pi \pi $ at a beam momentum of
1642 MeV/c.
Histograms show fits to data.}
\end{center}
\end{figure}

In our earlier publication on the present data in the year
2000, a sizable $a_0(980)\sigma$ amplitude was fitted.
It accounted for $\sim 15\%$ of the $\eta 3\pi^0$ cross section, see
Fig. 2(a) of Ref. \cite {eta3pi}.
It was fitted using the 1993 parametrisation of the $\sigma$
amplitude by Zou and Bugg \cite {zoubugg}.
Since then, the understanding of the $\sigma$ amplitude has
improved greatly through the work of Caprini et al.
\cite {Caprini}, using the Roy equations.
That work fits the $\pi\pi$ scattering length to
$(0.220 \pm 0.005) m_\pi^{-1}$ and alters the $s$-dependence
of the $\pi\pi$ amplitude below 600 MeV quite significantly.
In addition, there is now evidence for significant coupling
of the $\sigma$ to $KK$: $g^2_{KK}(\sigma)/g^2_{\pi\pi}
(\sigma) = 0.6 \pm 0.15$ \cite {buggKK}.
We now use Eqs. (1)-(11) of \cite {sigpole} for the $\sigma$;
they fit accurately the predictions of Colangelo et al. up
to 1 GeV and fit better the inelasticity required above the
$KK$ threshold.
The additional structure in the $\pi\pi$ amplitude is modest,
but enough that the evidence for the $a_0(980)\sigma$
amplitude in present data almost disappears.
The remaining signal is barely a 2 standard deviation effect.
It is now omitted and systematic errors covering the possible
signal will be included in branching ratios discussed below
in subsection 4.2.
The earlier $a_0(980)\sigma$ signal was perturbing the fit
via interferences with $a_0(980)$ signals from decays of
$\eta_2(1645)$, $\eta_2(1870)$ and $\eta_2(2030)$.
A consequence is that there are now rather large
changes to the branching ratios of the $\eta_2(1870)$ and
$\eta_2(2030)$ to $a_0(980)\pi$.
The basic difficulty here is that the earlier broad $\sigma$
amplitude had little structure and gave a rather flexible
fit to the data.
The branching ratios of $\eta_2(1645)$ and $\eta_2(1870)$ to
$f_2(1270)\eta$ and $a_2(1320)\pi$ are more robust and change
little.

The $\sigma \to \pi \pi$ amplitude is needed only up to $\sim 1100$
MeV.
Its form for elastic scattering is known quite precisely.
In elastic scattering, it is parametrised in the form $N(s)/D(s)$,
where the numerator contains an Adler zero just below threshold,
making the amplitude weak at low momenta.
In some production processes with large momentum transfers, e.g.
$J/\Psi \to \omega \sigma$, the numerator needs to be replaced with a
constant in order to reproduce a broad peak in the mass range 450-500
MeV, produced by a pole in $D(s)$.
In the present data, there is no evidence for this behaviour, so the
amplitude is taken to be that of elastic scattering.

\section {Fits to data}
In order to demonstrate the existence of $\eta_2(1645)$,
$\eta_2(1870)$ and $\eta_2(2030)$, it is necessary to fit data at all
beam momenta and show that various selections of events require the
presence of all three, with consistent masses, widths and ratios of
decay amplitudes at all beam momenta.
The earliest publication studied data at just two beam momenta,
1200 and 1940 MeV/c \cite {Adomeit}.
It was immediately obvious that these two momenta required two
resonances $\eta _2(1645)$ and $\eta _2(1870)$, produced with
considerably different relative intensities at the two momenta.
The $\eta_2(1870)$ has a strong decay to $f_2(1270)\eta$ and
the $\eta_2(1645)$ does not.
The picture developed further when data at all nine beam momenta were
available.
A further $\eta _2(2030)$ was required, with a distinctive decay to
$[a_2\pi ]_{L=2}$ \cite {eta3pi}.
Here we shall not repeat this lengthy story, but refer the reader to
the original publications.
The conclusion from all beam momenta combined is that just these
three states are sufficient to fit all the data, with consistent
decay amplitudes at all beam momenta.
Here we simply illustrate the quality of fits to mass projections.

Fig. 4 shows mass projections for $3\pi$, $\pi \pi \eta$, $\pi \eta$
and $\pi\pi$ at a beam momentum of 1642 MeV/c.
Points with errors are data; fits are shown as histograms.
In (a), there is a high mass peak due to $\pi_2(1670)\eta$, $\pi _2 \to
f_2(1270)\pi$.
In remaining panels there are peaks due to $f_1(1285)$, $a_0(980)$
and $a_2(1320)$ and a shoulder due to $f_2(1270)$.

At this beam momentum, the $\pi \pi \eta$ mass spectrum does not
distinguish $\eta _2(1645)$, $\eta_2(1870)$ and $\eta_2(2030)$ cleanly.
It is necessary to select events in the mass range 1500--1750 MeV to
study properties of $\eta _2(1645)$.
Figs. 5(a) and (b) show $\pi \eta$ and $\pi \pi$ mass spectra for this
selection.
Panels (c) and (d) show  mass projections for the $\pi \pi \eta$
mass range 1775--1975 MeV, centred on $\eta _2(1870)$; Figs. 6(a) and
(b) show projections for the $\pi \pi \eta$ mass range 2.0--2.25 GeV.
Taken together with angular dependence in the data, these projections
constrain fits to $\eta_2(1645)$, $\eta _2(1870)$ and $\eta_2(2030)$ and
their individual decay modes.
\begin{figure}[htb]
\begin{center}
\vskip -10mm
\epsfig{file=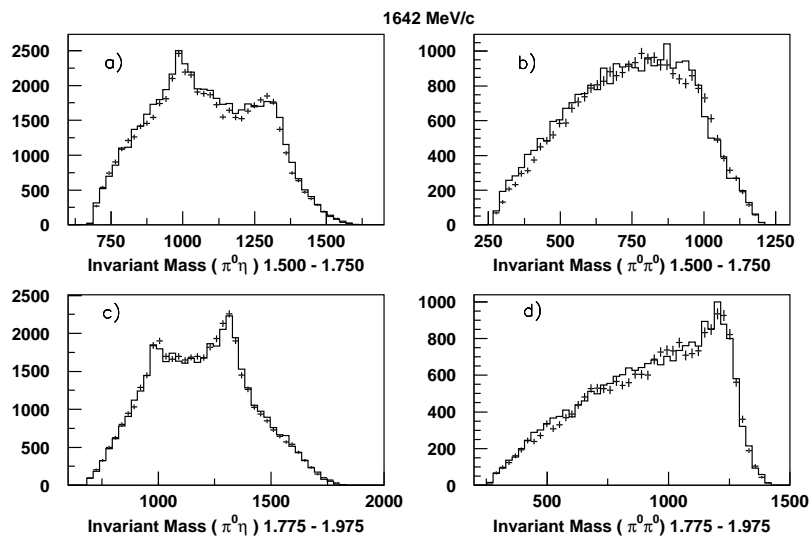,width=12cm}
\vskip -6mm
\caption {Mass spectra at 1642 MeV/c for (a) and (c) $\pi ^0 \eta$  in
two ranges of $M(\pi \pi \eta )$; (b) and (d) $\pi ^0\pi ^0$.
Histograms show fits to data.}
\end{center}
\end{figure}
\begin{figure}[htb]
\begin{center}
\vskip -10mm
\epsfig{file=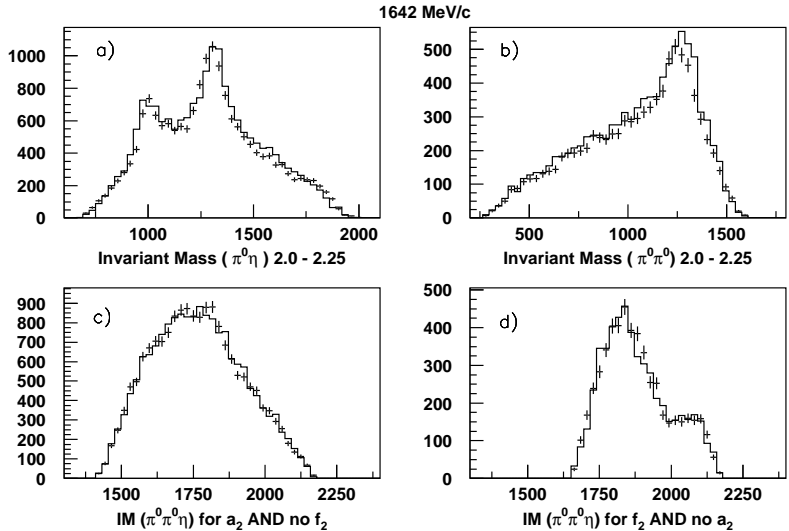,width=12cm}
\vskip -6mm
\caption {Mass spectra at 1642 MeV/c for
(a) and (b) $\pi ^0\eta$  and $\pi ^0\pi ^0$ in the
$M(\pi \pi \eta)$  mass range 2.0--2.25 GeV;
(c) the $\pi \pi \eta$ mass spectrum selecting $a_2$ and
vetoing $f_2(1270)$ (d) the converse selection.
Histograms show fits to data.}
\end{center}
\end{figure}

In order to display more clearly the $\eta _2(1870)$, Figs. 6(c) and (d)
shows mass projections which select $a_2(1320)$ and reject
$f_2(1270)$ or vice versa.
In (c), $a_2\pi$ events are selected with
$M(\pi \eta)$ in the mass range $1318 \pm 55$ MeV;
events with $\pi ^0\pi ^0$ in the mass range 1.0--1.455 GeV are
rejected.
In (d), $f_2(1270)$ is selected in the mass range 1.1--1.37 GeV
and events containing $\pi \eta$ in the mass range 1.15--1.435 GeV
are vetoed.
The strong $f_2(1870)$ peak in $f_2(1270)\eta$ is clearly visible
in Fig. 6(d).
In Fig. 6(c), one sees the combined $\eta\pi\pi $ mass spectrum
from $\eta _2(1645) $  and $\eta _2 (1870) \to a_2\pi$.
Similar cuts in data selection illustrate the presence of
$\eta_2(2030) \to [a_2\pi]_{L=2}$.
Fits to data are of similar quality at all beam momenta;
further examples were shown in Ref. \cite {Adomeit}.

In order to achieve good fits to data, the eleven channels of
Section 1 need to be sub-divided to include separate channels for
(i) $\eta_2(1645) \to a_2\pi $ and $a_0\pi$,
(ii) $\eta_2(1870) \to f_2(1270)\eta$, $a_2\pi$ and $a_0\pi$,
(iii) $\eta_2(2030) \to [a_2\pi]_{L=2} $, $[a_2\pi]_{L=0}$,
$a_0\pi$ and $f_2(1270)\eta$,
(iv) $f_2(1910) \to [a_2\pi]_{L=1} $ and $[f_2(1270)\eta]_{L=1}$,
(v)  $f_2(2001) \to [a_2\pi]_{L=3}$ only,
(vi) $f_2(2240) \to [a_2\pi]_{L=1,3}$ and $[f_2(1270)\pi]_{L=1,3}$.
At each beam momentum, phases for all these channels are fitted
freely.

In earlier work, there was some evidence for
$\eta_2(1870) \to [\eta \sigma]_{L=2}$.
There remains a small improvement in log likelihood when this
channel is included.
This improvement is however typically 30, which is less than for
almost all other channels.
Furthermore, its inclusion increases errors on other channels,
i.e. fits become less stable including it.
The problem is that there is no narrow signature of the
$\sigma \to \pi \pi$ S-wave, so it tends to absorb any noise in
the data.
This decay is now omitted.

\begin{table}[htb]
\begin {center}
\begin{tabular}{cccccccc}
\hline
Momentum(MeV/c) & 1050& 1200& 1350 & 1525& 1642 & 1800 & 1940 \\
Events          &15709&35127&26379&21339 &25394 &28200 &31388 \\\hline
channel \\
$f_2a_0$        &  163 & 344 &  408 & 471 &  409 & 357  &  281 \\
$a_2\sigma$     &  195 & 353 &  455 & 268 &  316 & 179  &  249 \\
$\pi_2(1670)\pi$&  -  & 337 &  384 & 381 &  655 & 790  &  390 \\
$\eta_2(1645)\to a_2\pi$& 98  &162 & 240 &  226 & 367 & 295  & 445 \\
$\eta_2(1645)\to a_0\pi$& 8   & 19 & 97  &  57  & 100 & 52   & 44 \\
$\eta_2(1870)\to f_2\eta$&206 &227 & 169 &  134 & 149 & 285   &324 \\
$\eta_2(1870)\to a_2\pi $&39  &127 & 66  &  35  & 57  & 21   & 80 \\
$\eta_2(1870)\to a_0\pi $&26  & 35 & 23  &  25  &  9  & 12  & 23 \\
$\eta_2(2030)\to [a_2\pi]_{L=2}$&161&294 & 336  & 367 & 440 & 438&203\\
$\eta_2(2030)\to [a_2\pi]_{L=0}$&9 & 17  & 21   &  6  & 9   & 12 &1\\
$\eta_2(2030)\to  f_2\eta      $&42& 101 & 150  & 124 & 109 &111 &39\\
$\eta_2(2030)\to  a_0\pi      $&77& 88  & 45   &  55 & 90   &14  &14\\
$f_2(1910)\to [a_2\pi]_{L=1}$& 359 & 644 & 330 & 223 & 269  & 179&96\\
$f_2(1910)\to [f_2\eta]_{L=1}  $& 17  & 27  & 37  & 49  & 112 & 54&71\\
$f_2(2001)\to [a_2\pi]_{L=3}$&22 & 20  & 63  & 152 & 279  & 88 &27\\
$f_2(2240)\to [a_2\pi]_{L=1,3}$& - & - & -&  -  &  -   & 432 &995\\
$f_2(2240)\to [f_2\eta]_{L=1,3}$&- & - &  - &  - &  -   & 19 &28\\
\hline
\end{tabular}
\caption{Changes in log likelihood when
each channel is removed from the fit and others are re-optimised.}
\end {center}
\end{table}
\begin{figure}
\begin{center}
\vskip -12mm
\epsfig{file=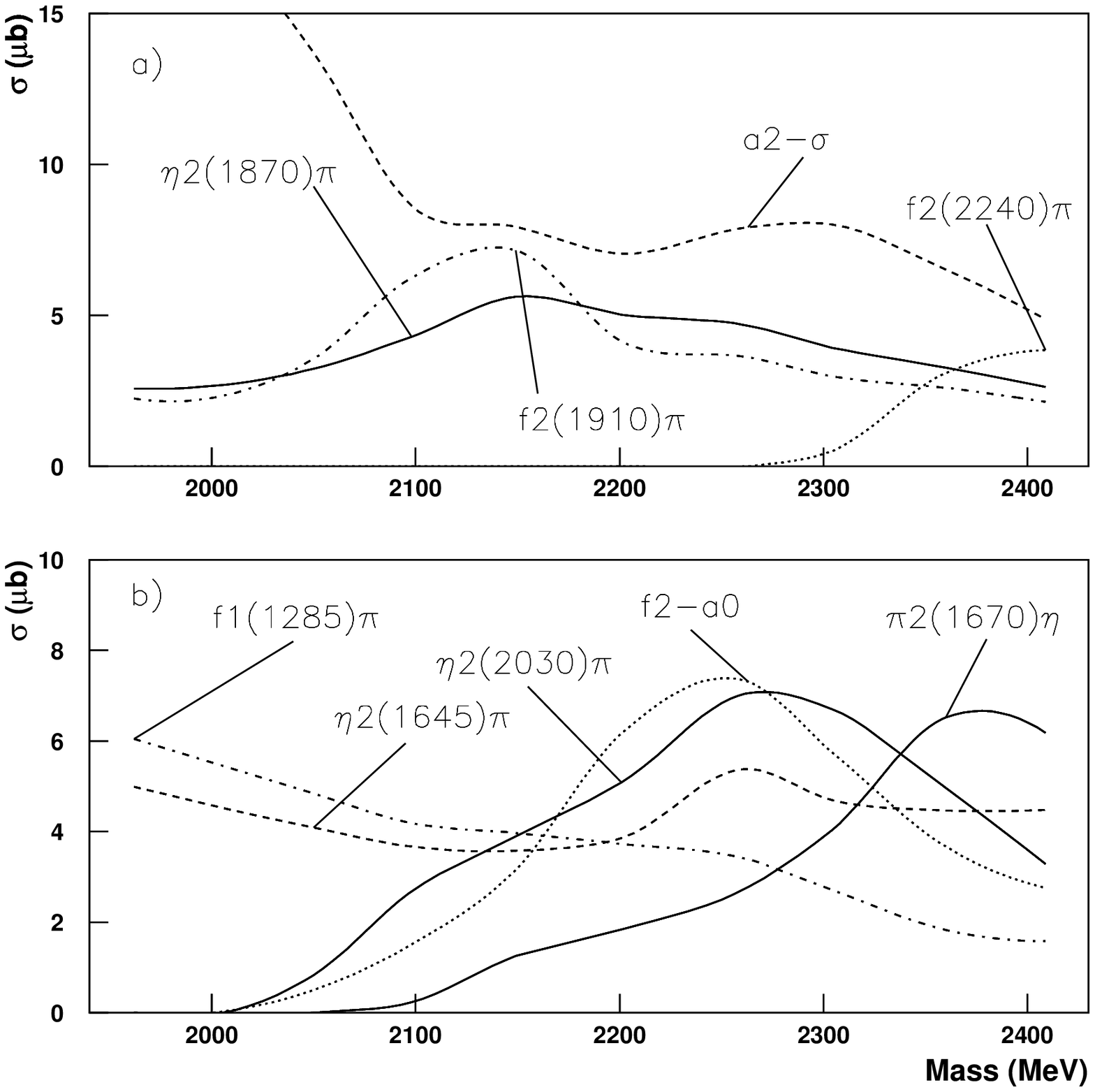,width=10cm}
\vskip -6mm
\caption {Cross sections for each final state. Smooth curves are
drawn through cross sections for each channel.}
\end{center}
\end{figure}

\section {Results}
Table 1 shows changes in log likelihood when each channel is omitted
from the fit and all other channels are re-optimised.
One sees immediately the significance level of each channel and its
dependence on beam momentum.

We now wish to draw conclusions from peaks observed in some channels
of data.
Fig. 7 shows cross sections v. beam momentum.
The absolute normalisation is taken from cross sections determined
in Ref. \cite {eta3pi} for the whole $\eta 3\pi ^0$ data; they
are uncorrected for branching fractions of $\eta$ and $\pi ^0 \to
\gamma \gamma$ and therefore correspond directly to the number of
events collected.
The integrated cross section varies little with beam momentum, with a
small $(10\%)$ enhancement near 2270 MeV; it is shown in Fig. 2 of
Ref. [3].
On Fig. 7, smooth curves for individual channels (2)--(12) are drawn
through the data within one standard deviation.
Errors are typically  $5-10\%$ and some examples will be
displayed with errors in Fig. 8.
Cross sections for several channels expand at low beam momenta 900 and
600 MeV/c to follow the $1/v$ dependence of the total cross section,
where $v$ is the beam velocity in the centre of mass.

In Fig. 7(b), there is a strong peak in the channel $f_2a_0$ near
2250 MeV and a broader peak in $\eta _2(2030)\pi$ at slightly higher
mass.
There is a possible peak in $\eta _2(1645)\pi$ in the same mass range.
There is also a peak in $f_2(1910)\pi$ near 2150 MeV.
The channel $\pi_2(1670)\eta$ peaks strongly at high mass.
There is evidence for production of $f_2(2240)\pi$ at the highest two
beam momenta.
Other channels show only weak structure, except that $a_2\sigma$ peaks
at low masses.

The question arises how to interpret the peaks.
Are they due to resonances?
Here it is necessary to take care over details in the formulae.
The amplitude for a resonance such as $a_2(2255) \to f_2(1270)a_0$
is
\begin {equation}
A  = \sqrt {\rho (\bar pp, s)\rho (f_2a_0, s)}/[k (M^2 - s - iM\Gamma
_{tot})].
\end {equation}
The factor $k$ in the denominator is the momentum of the $\bar p$ in
the $\bar pp$ centre of mass; this is the
flux factor for the incident beam.
The phase space factor for $\rho (\bar pp, s)$ in the numerator is
$k/\sqrt {s}$, multiplied by a centrifugal barrier factor.
We choose to make comparisons with data by accounting explicitly for
the intensity factors $k/\sqrt {s}$ for $\bar pp$ and $1/k^2$ from
the flux factor; the two together give a factor $1/k\sqrt {s} = 1/v$,
where $v$ is relativistic velocity.
We multiply cross sections of Fig. 7 by a factor $v/v_0$, where $v_0$
is evaluated at 2410 MeV, the highest data point. Remaining centrifugal
barrier factors and form factors are less certain and are modelled in
the fit to results.

\begin{figure}[htb]
\begin{center}
\vskip -12mm
\epsfig{file=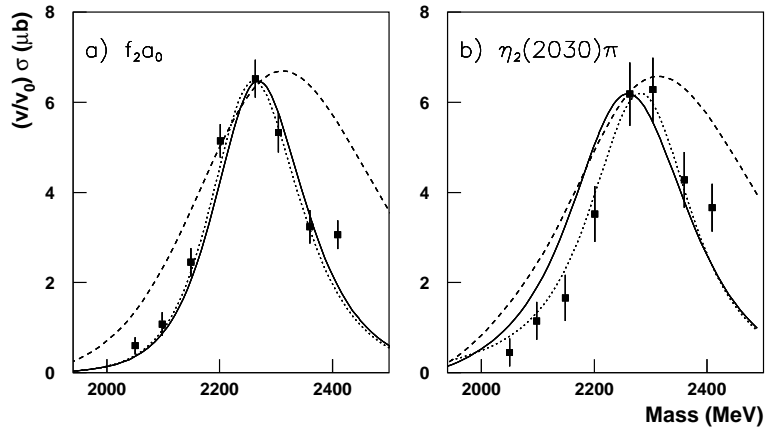,width=12cm}
\vskip -6mm
\caption {Production of  (a) $f_2(1270)a_0$ compared to
phase space for S-wave production (dashed curve), a fit
to $f_2(2255)$ (full curve) and a fit with parameters of this state
optimised (dotted curve); (b) $\eta_2(2030)\pi$ compared to
phase space for P-state production (dashed curve), a
fit using the line-shape of $\pi_2(2245)$ (full curve) and a fit with
parameters of this state optimised (dotted).}
\end{center}
\end{figure}
Fig. 8 shows data for $(v/v_0)\sigma$.
For $f_2a_0$ in Fig. 8(a), there is a peak closely resembling the
$a_2(2255)$, reported in the earlier combined analysis of
$I=1$, $C =+1$ CBAR data with $M=2255 \pm  20$ MeV and $\Gamma =
230 \pm 15$ MeV; it was observed as a clear peak in data for
$\bar pp \to f_2(1270)\pi$, see panel $(\ell )$ of Fig. 38 of Ref.
\cite {review}.
It was observed in both $^3F_2$ and $^3P_2$ decays to $f_2(1270)\pi$,
with an amplitude ratio $-2.13 \pm 0.20$ favouring coupling of
$\bar pp$ to $^3F_2$.
The dashed curve of Fig. 8(a) shows the remaining phase space
factor calculated with this ratio for $^3F_2$ and $^3P_2$
production and with S-wave decay to $f_2a_0$.
The rise of this curve with mass fails to fit the data.
Any other ratio of $^3P_2$ and $^3F_2$ also fails to fit the data.
Adding P-state $f_2a_0$ decays makes the dashed curve peak even higher
in mass.

The full curve shows the result of multiplying the dashed curve by the
line-shape of $a_2(2255)$, assuming a Breit-Wigner resonance of
constant width.
It is remarkably close to the data.
We regard this as further confirmation of the $a_2(2255)$.
A small improvement is possible by reducing the mass by 10 MeV to
2245 MeV and decreasing the width to 225 MeV, but these changes
are within the errors of the earlier determination and also
within errors of present data.

Fig. 8(b) shows results for the channel $\eta _2(2030)\pi$.
Production of this final state is dominantly ($\sim 74\%$) with
$\bar pp$ helicity 0, suggesting production via
$\bar pp \to \pi_2(2245)$:
\begin {eqnarray}
\bar pp \to \pi _2(2245) &\to& [\eta _2(2030)\pi ]_{\ell=1} \\
\eta_2(2030) &\to& [a_2\pi ]_{L=2}\\
a_2(1320) &\to& [\eta \pi]_{L=2}.
\end {eqnarray}
The full amplitude $D$ for this process is rather distinctive.
Suppose it is written in fixed axes in the $\bar pp$ centre of mass.
The initial state is spin singlet, helicity 0.
Suppose the spectator pion of Eq. (18) is produced at angle $\theta$ of
Fig. 2 and with azimuthal angle $\phi$ around the beam direction.
Let the momentum  of the decay pion in Eq. (18) be $q$.
Let the decay angle of the $a_2(1320)$ in Eq. (20) after the Wick
rotation be $\alpha$, with azimuthal angle $\beta$, using the same axes
as for Eq. 18.
Then
\begin {eqnarray} D &\propto& \sqrt \frac{1}{2} q B_1(q)
\sin \theta \sin \alpha \cos \alpha (e^{i(\phi - \beta )} -
e^{-i(\phi - \beta )}) \\
  &=& i \sqrt {2} q B_1(q) \sin \theta \sin \alpha \cos \alpha
\sin (\phi - \beta ).
\end {eqnarray}
Here $B_1(q)$ is the centrifugal barrier factor for $L=1$ decay.
The amplitude for the decay $a_2(2030) \to f_2(1270)\eta$ has an
identical form.
Amplitudes for decays of $a_2(2030) \to [a_0(980)\pi]_{L=2}$
take a similar form with angles $\alpha$ and $\beta$ those of the
decay pion.
The amplitude for decays to $a_2(2030) \to [a_2(1320)\pi]_{L=2}$
involves a combination of spin 2 of the $a_2(1320)$ with $L=2$ to
make spin 2 of the $a_2(2030)$.
This amplitude may be constructed along the same lines using
Clebsch-Gordan coefficients or using tensor algebra.
Because of the additional angular dependence on $L=2$
in the decay to $[a_2(1320)\pi ]_{L=2}$, it is particulary distinctive
and stands out clearly in the data.

From earlier work, the parameters of $\pi_2(2245)$ have sizable errors,
$M = 2245 \pm 60$ MeV, $\Gamma = 320 ^{+100}_{-40}$ MeV \cite {I1}.
The full curve of Fig. 8(b) shows a fit with these parameters;
the width is clearly too large and the mass somewhat too low.

Production of $\eta_2(2030)\pi$ is $(26 \pm 4)\%$ in intensity from
$\bar pp$ helicity 1 at the peak; this helicity 1 component may be
accomodated with $a_2(2255) \to $ S-wave $\eta _2(2030)\pi$.
Note that $a_2(2255)$ cannot contribute to
$[\eta _2(2030)\pi]_{\ell=1}$.
Also $[\eta _2(2030)\pi]_{\ell=2}$
has a strong variation with mass and will be strongly suppressed by
the $\ell = 2 $ centrifugal barrier.
The dotted curve of Fig. 8(b) is obtained by
adjusting the mass and width of $\pi _2$ to $M=2285 \pm 20(stat)
\pm 25(syst)$ MeV,
$\Gamma = 250 \pm 20(stat) \pm 25(syst)$ MeV and including $26\%$
of the intensity via $a_2(2255)$.

The data determine relative phases of strong channels with errors
of $\sim \pm 15^\circ$ at every momentum.
The phase variation of $f_2a_0$ and $a_2(2030)\pi$ channels agree
within these errors.
If $f_2a_0$ is resonant, then so is $a_2(2030)\pi$ and this
can only be explained in terms of some combination of
$\pi _2(2245)$ and $a_2(2255)$.
There are further triplet states $a_3(2275)$ and $a_1(2245)$ in this
vicinity, but they can only contribute to decays to
$[\eta_2(2030)\pi]_{\ell=2}$; this is inconsistent with the
observed line-shape of the peak and can only make a small
contribution to the data.
Systematic errors assigned to the mass and width of $\pi _2(2245)$
cover any weak contribution from this source.

\subsection {Other channels}
The production of $\eta _2(1875)\pi$ is distinctively different
to that of $\eta_2(2030)\pi$.
Production from the initial $\bar pp$ system with helicity $\pm 1$
is strongly dominant, requiring an initial triplet state.
However, no conclusion can be drawn from the slowly varying
cross section in Fig. 7(a).
A minor correction to Ref. \cite {eta3pi} is that the $\eta
_2(1875)\pi$ signal plotted there in Fig. 2(a) was multiplied by
a factor 2 to make it clearly visible; the intensity recorded there
is close to the present analysis.

Production of $f_2(1910)\pi$ is roughly equal from initial states
with helicity 0 and 1, but shows significant slow variation with mass.
This is possible from initial triplet states.
The peak near 2150 MeV may indicate production from the initial state
$a_2(2175)$.

Production of $\pi_2(1670)\eta$ is dominantly via $\bar pp$ helicity
1, (i.e. spin triplet) but again fluctuates smoothly with mass by more
than a factor 2.
No firm conclusion can be drawn from the variation of cross
section with mass.

A final point is that data are available at 900 MeV/c and were included
in the earlier analysis \cite {eta3pi}.
However, there are small cross sections for several channels and these
are difficult to determine with confidence.
This beam momentum has been studied, but is discarded from the
present analysis because of substantial systematic errors in weak
channels.
There are also low statistics at 600 MeV/c, and this momentum is
omitted for the same reason.

\subsection {Branching fractions}
Fortunately, the branching fractions of $\eta_2(1645)$, $\eta _2(1870)$
and $\eta_2(2030)$ are not sensitive to the question whether or not
their production goes directly via resonances in the $\bar pp$ channel.
If such resonances are involved, all decay modes of these
channels pick up the same phase variation from the production process.
The dominant decay of $\eta_2(2030)$ is to $[a_2\pi]_{L=2}$;
that for $\eta _2(1870)$ is to $f_2(1270)\eta$ and that for
$\eta _2(1645)$ is to $a_2\pi$.
Branching fractions will be quoted with respect to these dominant
channels.
Some of the smaller branching fractions have changed significantly
since the analysis of the year 2000.

A general comment is that data determine ratios of amplitudes
rather than ratios of intensites.
That is, log likelihood has a parabolic minimum as a function of
the amplitude ratio.
The procedure is therefore to fit data using, for example,  the
magnitude of the $\eta_2(1645) \to a_2\pi$ amplitude as one variable
and the ratio of amplitudes
$\eta _2(1645) \to a_0\pi /\eta_2(1645) \to a_2\pi$ as the second.
This makes it easy to investigate fluctations with beam momentum.
The average of each ratio of amplitudes is found by weighting
values at each beam momentum with the total number of $a_2(1645)$
events summed over both channels.
The same procedure is used for $\eta_2(1870)$, $\eta _2(2030)$ and
$f_2\pi$ channels like $f_2(1910)\pi$.
This gives more stable results than trying to determine
errors of ratios at each momentum and using them in the weighting
procedure; errors fluctuate somewhat from momentum to
momentum because of large errors for phases fitted to weak channels.
Having found the average ratios of amplitudes, a second pass
is made through the fit, fixing these ratios at all momenta.
Then the ratio of intensities is obtained from the total number
of events fitted to each channel, integrated over all beam momenta.

In evaluating branching ratios, it is necessary to make use
of Eqs. (13)-(15).
Here there is a dilemma.
It is convenient to parametrise the Breit-Wigner denominator as far
as possible with constant widths for each decay channel, as is
conventional in the Particle Data Tables.
Our procedure is to modify the numerator of Eq. (15) to
$\Gamma_2F(s)$, so as to agree with the denominator.
We then evaluate branching ratios at each beam momentum, and average
over momenta to determine $\Gamma_1/\Gamma_2$.
This procedure converges within errors after one iteration.
The same procedure is used to evaluate the effects of $L=2$ centrifugal
barriers on widths for $\eta_2(2030) \to [a_2\pi]_L=2$ and
$\eta_2(2030) \to [a_0\pi]_{L=2}$.

The acceptance for strong channels $f_2(1270)\eta$ and
$[a_2\pi]_{L=2}$ varies rather strongly with beam momentum.
Numerical results are shown in Table 2, normalised to 1 at the
highest beam momentum 1940 MeV/c; they are evaluated using the
Monte Carlo simulation of the detector.
One sees a large variation of the intensity ratio $R$ for some
channels.
This variation almost disappears at 1940 MeV/c, where all decay
channels are nearly fully open.
We shall tabulate branching ratios $R$ corrected to this momentum.
Above this momentum, results may be affected by errors in the Fermi
function adopted in Eq. (14), so this is close to the optimum
compromise.
Errors in branching ratios due to these uncertainties are included
in errors quoted in Table 3 below.
We regard this procedure as an improvment on the work of Ref. [3],
where branching ratios were evaluated purely from geometric
acceptance without the effects of centrifugal barriers or the form
factor $FF(s)$ of Eq. (13).
\begin{table}[htb]
\begin {center}
\begin{tabular}{ccccccccc}
\hline
momentum(MeV/c): & & 1050 & 1200 & 1350 & 1525 & 1642 & 1800 &
1940\\
Resonance & Ratio \\\hline
$\eta_2(1645)$ &$a_0\pi/a_2\pi$  & 0.955 & 0.974 & 0.986 & 0.990 &
0.993 & 0.997 & 1.0 \\
$\eta_2(1870)$ &$a_2\pi/f_2\eta$ & 2.32 & 1.80 & 1.50 & 1.31 &
1.13 & 1.06 & 1.0 \\
             & $a_0\pi/f_2\eta$   & 2.01 & 1.51 & 1.13 & 1.04 &
             1.01 & 1.001 & 1.0 \\
$\eta _2(2030)$ & $[a_2\pi]_{L=0}/[a_2\pi]_{L=2}$ & 1.78 & 1.57 &
1.34 & 1.19 & 1.08 & 1.04 & 1.0\\
                & $a_0\pi/[a_2\pi]_{L=2}$ & 1.57 & 1.43 & 1.26 &
1.15 & 1.09 & 1.04 & 1.0 \\
               & $f_2\eta/[a_2\pi]_{L=2}$ &  0.98 & 1.00 & 1.01 & 1.02
& 1.01 & 1.01 & 1.0 \\

$f_2(1910)$ & $f_2\eta/a_2\pi$ & 0.738 & 0.803 & 0.871 & 0.917 & 0.948
               & 0.976 & 1.0\\\hline
\end{tabular}
\caption{Variation of acceptance (including form factors) with beam momentum}
\end {center}
\end{table}

Coming to technicalities, it will be necessary to correct the number of
observed $a_2(1320)\pi$ and $f_2(1270)\eta$ events in $\pi ^0
\pi^0 \eta $ for unobserved decays.
The $f_2(1270)$ has a branching ratio of
$0.848/3$ to $\pi ^0 \pi ^0$ \cite {PDG}.
The amplitude for the $I=1$ component of $\bar pp \to \pi X$,
$X \to f_2\eta$ is given by
\begin {eqnarray}
A(I=1) &=& \pi ^0_1 X^0_{23} + \pi ^0_2 X^0_{31} + \pi ^0_3 X^0_{32} \\
       &\to&\sqrt{1/3}[\pi ^0_1 (\pi ^0_2 \pi ^0_3)\eta
+ \pi ^0_2 (\pi ^0_3\pi ^0_1)\eta + \pi ^0_3(\pi ^0_1\pi ^0_2)\eta].
\end {eqnarray}
In (23), $X$ stands  for the amplitude of $X \to f_2\eta$; the brackets
in the last line identify pions coming from $f_2(1270)$.
For $X \to a_2\pi$, the decay amplitude of $X$ is $\sqrt {1/3}(\pi
^+ a_2^- -\pi ^0 a_2^0 + \pi ^- a_2^+)$, where the minus sign for $\pi
^0 a_2^0$ can be absorbed into the fitted phase for this channel.
In the $\pi ^0 a_2^0$ final state, what is actually observed is
\begin {equation}
B(I=1) =\sqrt{1/6} \left( \pi ^0_1 [(\pi ^0_2 \eta)\pi^0_3 +
(\pi ^0_3\eta)\pi ^0_2] + \pi ^0_2 [(\pi ^0_3\eta)\pi^0_1 + (\pi
^0_1\eta)\pi ^0_3]
+ \pi ^0_1 [(\pi ^0_2\eta)\pi^0_3 + (\pi ^0_3\eta )\pi
^0_2] \right) ,
\end {equation}
i.e. six combinations. The coherent sum of all
combinations is fitted to the data.

Table 3 shows in column 3 branching fractions from the
previous analysis \cite {eta3pi} for comparison purposes.
The next column lists what is fitted now.
The final column corrects this for all charge states
and for the branching fractions of $a_2(1320) \to \pi \eta$ $(14.5\%)$
and $a_0(980)\pi$ to $\pi \eta$.
This last branching fraction is taken to be the value used
by WA102, $(86\%)$ for easy comparison with their results.
This is close to the value adopted by the PDG \cite {PDG}.
If their value is adopted, $a_0(980)$ branching fractions increase
by a factor $1.015 \pm 0.021$.
Results in columns 4 and 5 supercede the earlier results.

Branching ratios of decays to $a_0(980)\pi$ final states are unstable.
They depend somewhat on whether or not the $a_0(980)\sigma$ channel is
included in the fit.
The basic difficulty is that the $\sigma$ amplitude is broad and gives rise
to interferences all over the 4-body phase space.
The $a_0\sigma$ channel is therefore not well determined.
In the absence of definite evidence that it is needed, we omit it.
All $a_0$ signals are weak and their phases with respect to
dominant decays have quite large errors.
There are also strong correlations between couplings of
$\eta_2(1645)$, $\eta _2(1870)$ and $\eta _2(2030)$ to $a_0\pi$.
In the earlier analysis, $a_0\pi$ decays interfered with $a_0\sigma$,
giving apparently small but unreliable errors.

From the present analysis, the ratio for $\eta _2(1645)$ lies close to
the WA102 result $0.077 \pm 0.016$; the $a_0(980)$ signal is clearly
visible as a peak in their raw data \cite {WA102A}.
For this reason, the weighted mean of these two values has been
adopted and fixed for $\eta _2(1645)$.
This helps stabilise fits to $a_0\pi$.
However, there are large correlations between $\eta _2(1870)$ and
$\eta_2(2030)$ decays to $a_0(980)\pi$.
Together with uncertainties due to possible contributions from
$a_0(980)\sigma$, the result is that branching fractions of
$\eta_2(1870) \to a_0\pi$ and $\eta_2(2030)\pi$ can both vary
freely over the range 0.1 to 0.85 in present data.
Errors on values given in Ref. \cite {eta3pi}
for these two channels need to be increased substantially to take
account of these systematic errors.
New estimates are given in Table 3 .
These uncertainties are not significantly correlated with branching
fractions to $f_2(1270)\eta$ and $a_2\pi$.
\begin{table}[htb]
\begin {center}
\begin{tabular}{ccccc}
\hline
Resonance & Ratio & Ref. [3] & Present data & Corrected \\\hline
$\eta _2(1645)$ & $BR(a_0\pi)/BR(a_2\pi)$  & $0.36 \pm 0.12$ &
                  $0.38 \pm 0.13$          & $0.074 \pm 0.025$ \\
$\eta _2(1870)$ & $BR(a_2\pi)/BR(f_2\eta)$ & $0.22 \pm 0.03$ &
                  $0.28 \pm 0.07$         & $1.60 \pm 0.40$ \\
$\eta _2(1870)$ & $BR(a_0\pi)/BR(f_2\eta)$ & $0.85 \pm 0.05$ &
                  $0.48 \pm 0.45$         & $0.48 \pm 0.45$ \\
$\eta _2(2030)$ & $BR(a_0\pi)/BR([a_2\pi]_{L=2})$ &  $0.37 \pm 0.08$ &
                  $0.59 \pm 0.50$         & $0.10 \pm 0.08$ \\
$\eta _2(2030)$ & $BR(f_2\eta)/BR([a_2\pi]_{L=2})$ & $0.43 \pm 0.15$ &
                  $0.75 \pm 0.34$         & $0.13 \pm 0.06$ \\
$\eta _2(2030)$ & $BR[a_2\pi]_{L=0}/BR([a_2\pi]_{L=2})$ & -
                & $0.05 \pm 0.03$         & $0.05 \pm 0.03$ \\
$f _2(1910)$    & $BR(f_2\eta)/BR(a_2\pi)$ &  -
                & $0.54 \pm 0.29$         & $0.09 \pm 0.05$ \\\hline
\end{tabular}
\caption{Branching ratios; column 4 shows values for present data and
column 5 shows values corrected for all charges and all decay modes of
$f_2$, $a_2$ and $a_0$.}
\end {center}
\end{table}
Branching ratios between $f_2(1270)\eta$ and $a_2(1320)\pi$ decay
modes are mostly quite stable, because these decays are strong.
The ratio of decays of $\eta _2(1870)$ to $f_2(1270)\eta$ and
$a_2(1320)\pi$ remains  stable.
So does the ratio of decays of $\eta _2(2030)$ to $f_2(1270)\eta$
and $[a_2\pi]_{L=2}$, because these are conspicuous decays.

A mistake has been located in the branching fraction of
$\eta _2(2030)$ to $[a_2\pi]_{L=0}$ reported in Ref. \cite {eta3pi}.
The corrected value is given in entry 6 of Table 3.
This decay is much weaker than that to $[a_2\pi]_{L=2}$.

\begin{table}[htb]
\begin {center}
\begin{tabular}{cc}
\hline
Beam momentum (MeV/c) & amplitude ratio $\eta _2(1870) \to a_2\pi/
\eta _2(1870) \to f_2\eta$ \\\hline
1050 & $0.383 \pm 0.067$ \\
1200 & $0.321  \pm 0.034$ \\
1350 & $0.219 \pm 0.050$ \\
1525 & $0.202 \pm 0.058$ \\
1642 & $0.320 \pm 0.050$ \\
1800 & $0.334 \pm 0.063$ \\
1940 & $0.308 \pm 0.065$ \\\hline
\end{tabular}
\caption{The ratio of amplitudes $\eta_2(1870) \to a_2(1320)\pi/
\eta_2(1870) \to f_2(1270)\eta$ at individual beam momenta.}
\end {center}
\end{table}
Table 4 gives the important branching ratio of amplitudes
$\eta_2(1870) \to a_2(1320)\pi/\eta_2(1870) \to f_2(1270)\eta$
at all beam momenta, in order to illustrate the stability.
We choose to take the simple mean $0.298$ over beam momenta, so as to
avoid bias from fortuitous fluctations in errors with beam momentum.
Fluctuations about the mean are above statistics by a factor 1.4.
The statistical error is increased to allow for this in Table 3.

In earlier work, the error on the mean was taken as the statistical
error divided by $\sqrt{N}$, where $N$ is the number of beam
momenta.
However, it is now clear that systematic errors are somewhat larger
than this.
The systematic errors are estimated from (i) variation of results with
the ingredients included in the fit, particularly the number
of interferences included between channels; (ii) variations with masses
and widths of $\eta_2(1645)$, $\eta _2(1870)$, $\eta _2(2030)$,
$f_2(1920)$, $f_2(2001)$ and $f_2(2240)$, (iii) possible
contribution from $f_2(2293)$, (iv) uncertainties in the background
from other final states.
Ultimately, systematic errors dominate for all branching ratios,
particularly for decays to the weak $a_0\pi$ channels.

\section {Fits to WA102 data}
The WA102 collaboration measured central production of $\pi \pi \eta$
and produced separate sets of data for $\eta \to \gamma \gamma$ and
$\pi^+\pi ^-\pi ^0$ \cite {WA102A}.
These data have been read from their graphs and refitted.

\begin{figure}[htb]
\begin{center}
\vskip -12mm
\epsfig{file=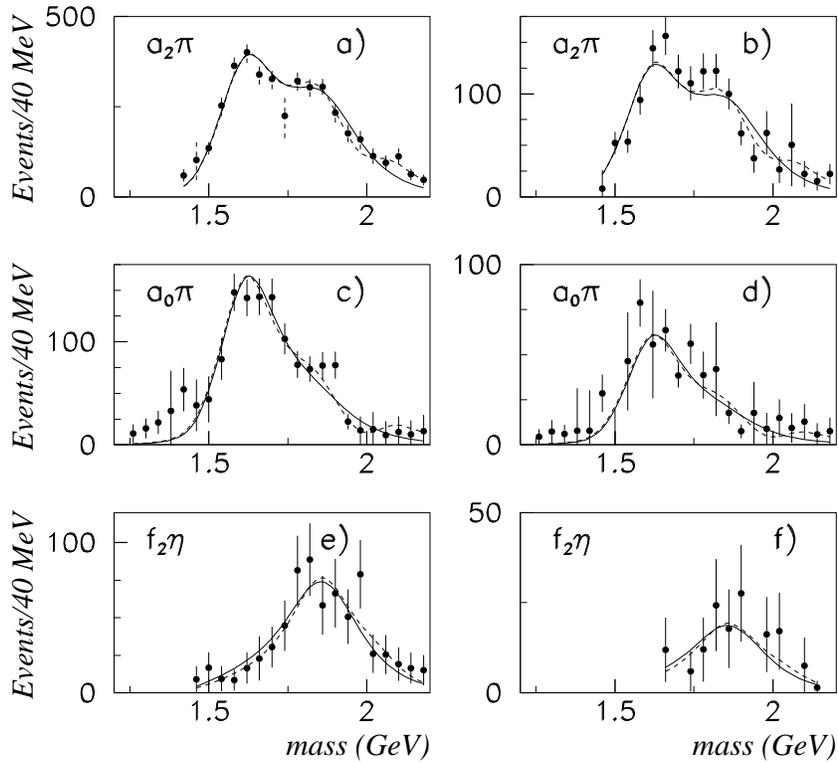,width=12cm}
\vskip -6mm
\caption {Fits to WA102 data for central production of $a_2\pi$,
$a_0\pi$ and $f_2(1270)\eta$ with $J^{PC} = 2^{-+}$.
The first column shows data for $\eta \to \gamma \gamma$ and the
second data for $\eta \to \pi ^+\pi ^-\pi^0$.
Full curves are fits without $\eta _2(2030)$ and dashed curves with it.}
\end{center}
\end{figure}Their approach was to use the K-matrix for $\eta _2(1645)$ and
$\eta _2(1870)$.
If a single amplitude is used in this approach, the amplitudes
for $\eta _2(1645)$ and $\eta _2(1870)$ each loop round the
Argand circle once; continuity of the amplitude then requires a
zero between them.
The dip due to this zero played a significant part in fitting the
data.

The basic assumption of the K-matrix approach is that resonances
combine in a production process in an identical way to elastic
scattering.
This is a pure assumption.
In elastic scattering, ingoing and outgoing waves for all
coupled channels must sum asymptotically to unit intensity.
However, in production processes considered here, the amplitude is only
a tiny fraction of the $\bar pp$ total cross section.
There is no obvious reason why 2-body unitarity should apply in the
same way as for elastic scattering.
The K-matrix approach has been tested on four sets of experimental
data in Ref. \cite {EU}; it failed seriously in every case.
If it is used, phases of $\eta _2(1645)$ and $\eta _2(1870)$ need
to be fitted freely, since final states may rescatter between one
another, generating phases which need to be fitted arbitrarily.
Our analysis therefore uses the isobar model.
This makes no assumption about the effects of unitarity and allows
separate phases for $\eta_2(1645)$ and $\eta _2(1870)$.

Refitted results are displayed on Fig. 9.
Data for $\eta \to \gamma \gamma$ and $\eta \to \pi ^+\pi ^-\pi
^0$ have been fitted simultaneously, including decays to all of
$a_0(980)\pi$, $f_2(1870)\eta$ and $a_2(1320)\pi$.
They have been fitted with and without $\eta _2(2030)$.
That component was not known at the time of the WA102 analysis.
Full curves on Fig. 9 show fits without $\eta _2(2030)$ and
dashed curves the fits including it.
With it, the total $\chi^2$ improves from 124.5 to 95.8.
The reason is obvious: it provides extra freedom in fitting small
defects in the mass region above 2 GeV.
This improvement in $\chi^2$ of 28.7 needs to be balanced against
the fact that there are three extra complex coupling constants for
the three channels of $\eta_2(2030)$, i.e. 6 extra fitting parameters.
The improvement in $\chi^2$ is 3.6 standard deviations.
It is debatable whether or not $\eta _2(2030)$ is really present.
Conclusions will be drawn here from fits without it; these fits are
more secure.

A significant point is that the K-matrix zero between the
1645 and 1870 MeV peaks made both of them narrower and pushed the
peaks apart.
In the WA102 fit, the mass  of $\eta _2(1645)$ was $1605 \pm 12$
MeV for $\eta \to \gamma \gamma$ data and $1619 \pm 11$ MeV for
$\eta \to 3\pi$.
These are to be compared with the CBAR determination of
$1645 \pm 6(stat) \pm 20(syst) $ MeV.
In the isobar model, the dip between the two resonances can be filled
in by interferences between them.
Table 5 shows fitted masses and widths for $\eta _2(1645)$ and
$\eta _2(1870)$ for two cases.
In the first, column 2, the parameters of $\eta _2(1870)$ are fitted
freely.

The width of the $\eta _2(1870)$ tends to run away to a large value.
In Fig. 9(d), the width of the $\eta_2(1870)$ peak in
decays to $f_2(1270)\eta$ is sensitive to scatter in the points near
the peak.
The second fit (column 3) is made adding to $\chi^2$ a contribution
given by the CBAR masses and widths with statistical and systematic
errors combined in quadrature.
This extra constraint stabilises the fit and gives mass and width
for $\eta _2(1870)$ closer to the CBAR values.
In our opinion, the third column is the more reliable, bearing
in mind that the addition of the $\eta _2(2030)$ increases the
uncertainties from WA102 data even further.

\begin{figure}[htb]
\begin{center}
\epsfig{file=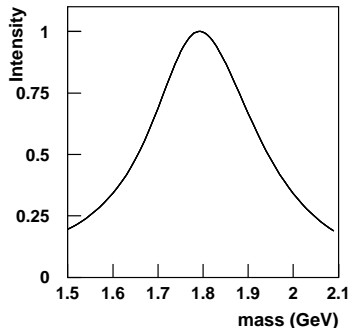,width=6cm}
\vskip -10mm
\caption {The line-shape of $\eta_2(1870)$ from the combined fit
to CBAR and WA102 data.}
\end{center}
\end{figure}

The explicit re-arrangement of Eq. (15) for $\eta _2(1870) \to
f_2(1270)\eta$ is
\begin {equation}  
f = \frac {[0.685 F(s) M\Gamma ]^{1/2}}
{[M^2 - s - i(0.672 + 0.685 F(s))M\Gamma]},
\end {equation}
where $M$ and $\Gamma$ refer to values for $\eta _2(1870)$ in the
last two lines of Table 6; $F(s)$ is given by Eq. (14).
The CBAR data are fitted using Eqs. (14) and (26).
Numerical values in the denominator are such that $|f|^2$
integrated over $s$ reproduces the branching fractions for
$a_2\pi$, $a_0\pi$ and $f_2(1270)\eta$ in Table 3.

The line-shape of $\eta_2(1870)$ is shown in Fig. 10 and is very
close to that of a Breit-Wigner resonance of constant width.
It peaks at 1792 MeV.
This is related to the opening of the $f_2(1270)\eta$ threshold.
The pole position is at $1798 \pm 20 - i(130 \pm 12)$ MeV.
If CBAR values are used instead, the imaginary part of the pole
position decreases to 109 MeV.
\begin{table}[htb]
\begin {center}
\begin{tabular}{cccc} \hline
 & Unconstrained fit & With CBAR constraint & CBAR values\\\hline
$\chi^2$      &  116.8         & 124.5 \\
$M(1645)$     &$1635 \pm 12$&$1630 \pm 9$&$1645\pm 6(stat)\pm 20(syst)$\\
$\Gamma(1645)$&$252 \pm 36$ &$225 \pm 16$&$200 \pm 5(stat)\pm 25(syst)$ \\
$M(1870)$     &$1833 \pm 20$&$1829 \pm 12$&
$1825\pm 5(stat)\pm 15(syst)$\\
$\Gamma(1870)$&$332 \pm 45$ &$293 \pm 24$&$221 \pm 20(stat)
^{+50}_{-35}(syst)$ \\\hline
\end{tabular} \caption{Masses and widths fitted to WA102 data without
any constraint (column 2) and with a penalty function given by errors on
CBAR masses and widths (column 3);
Column 4 shows CBAR values of masses and widths.}
\end {center}
\end{table}

The remaining question is whether there is really any significant
$a_2(1320)$ signal in WA102 data due to
$\eta _2(1870) \to a_2(1320)\pi$.
In a 1997 publication \cite {WA102B}, WA102 claimed to observe
$\eta _2(1645)\to a_2(1320)\pi$ in central production of $4\pi$,
with a small shoulder at high mass which could be $\eta _2(1870)$.
They reported a strong $2^+$ signal in $a_2(1320)\pi \to \rho \pi \pi$
at 1900 MeV and a broad $2^+$ peak in $f_2\pi \pi$ near 2000 MeV.
The integrated $2^+$ signal was considerably stronger than
that for $J^P = 2^{-+}$.
However, further data reported in the year 2000 on central production
of $4\pi$ were interpreted in terms of
$2^{-+} \to a_2(1320)\pi \to 4\pi$, produced only with $J_z = \pm 1$
\cite {WA102C}.
The $2^{++} \to f_2(1270)\pi \pi$ was found again but no $2^{++}
\to a_2(1320)\pi$.
They make no comment on why this change from the 1997 work occurs.
In central production via Pomeron exchange, there is
no obvious reason why $2^{++}$ should not be produced with $J_z = 0,$
$\pm 1$ and $\pm 2$.
Our view is that the data really need to be fitted with all allowed
values of $J_z$ for both $J^P = 2^+$ and $2^-$.
It would be valuable if the Compass collaboration could check this
point.

In the CBAR data analysed here, there is clear evidence for
$f_2(1910)$ and $f_2(2001)$ decaying to $a_2(1320)\pi$ and
$f_2(1270)\eta$.
The $a_2\pi$ decay dominates.
This is readily understood from the fact that the $L=1$ centrifugal
barrier inhibits decay to $f_2(1270)\eta$.
We suggest that the small bump claimed by WA102 in $a_2(1320)\pi$ at
1860 MeV is due to $f_2(1910)$.
We have fitted WA102 data using PDG parameters $M=1903$, $\Gamma =
196$ MeV for $f_2(1910)$ instead of $\eta _2(1870)$.
The fit, shown in Fig. 11, gives a slightly improved description of the
data (by 11 in $\chi^2$), but we are unable to go back to the original
data and check the $J^P$ analysis.
\begin{figure}[htb]
\begin{center}
\epsfig{file=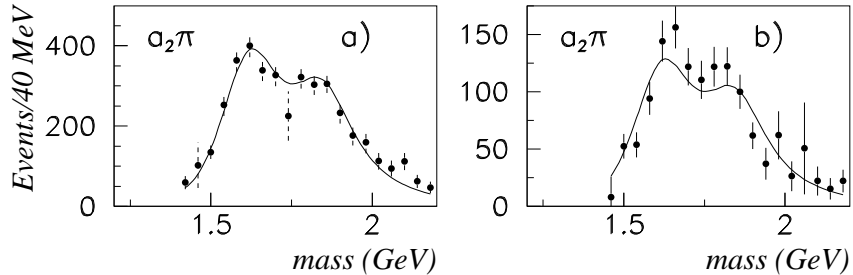,width=12cm}
\vskip -6mm
\caption {Fits to WA102 data with $\eta_2(1870) \to a_2\pi$
replaced by $f_2(1910) \to a_2\pi$: (a) $\eta \to \gamma \gamma$, (b)
$\eta \to \pi ^+\pi ^-\pi ^0$.}
\end{center}
\end{figure}

It remains an interesting question why $\eta _2(1870)$ has a fairly
large $f_2(1270)\eta$ S-wave decay.
The dispersive term $m(s)$ of Eq. (17) leads to attraction in this
channel near threshold \cite {Sync}.
It necessarily favours this decay mode.
There is an isospin partner $\pi _2(1880)$ for $\eta_2(1870)$.
It too has a strong decay mode to $a_2(1320)\eta$ \cite {E8523}.

\section {Evidence that the $\eta _2(1870)$ is resonant}
Several checks have been made that the $\eta _2(1870)$ has resonant
phase variation.
The first check is to remove the phase variation by
replacing the amplitude by its modulus and refitting the data.
At all beam momenta, this leads to a highly significant worsening
of log likelihood.
Column 2 of Table 5 shows the changes against beam momentum.
The definition of log likelihood is such that a change of 0.5 in
log likelihood should correspond to a one standard deviation change.
We have already remarked that fluctuations in Table 4 are a factor
1.4 above statistics.
A more extensive examination of fluctuations in branching ratios
shows they are in some cases up to a factor 2 above statistics, and
that has already been taken into account in errors quoted in Table 3.
Adopting this as a general rule leads to the conclusion that
changes listed in Table 3 can  be equated to changes in $\chi ^2$.
They average to 11.6 standard deviations per momentum, i.e.
28$\sigma$ in total.
\begin{table}[htb]
\begin {center}
\begin{tabular}{ccc}
\hline
Momentum (MeV/c) & (i) & (ii) \\\hline
1050 & 43 & 31 \\
1200 & 81 & 98 \\
1350 & 106 & 120 \\
1525 & 91 & 47 \\
1642 & 95 & 55 \\
1800 & 195 & 127 \\
1940 & 199 & 86 \\\hline
\end{tabular}
\caption{Changes in log likelihood when the amplitudes for
$\eta _2(1870)$  (i) are replaced by their modulus (no phase
variation), (ii) use a denominator $A - m(s) - iM\Gamma (s)$,
see text.}
\end {center}
\end{table}

The second check is that including the dispersive term of Eq. (16)
into the Breit-Wigner denominator has little effect on log
likelihood after minor alterations to fitted mass and width, well
within errors quoted in Table 6.
The dispersive term peaks at the $f_2(1270)\eta$ threshold, but
with a large full width of $\sim 300$ MeV.
Numerically, it is easily absorbed into small shifts of fitted
parameters.

There remains the possibility that the associated cusp in the real
part of the amplitude could explain the phase variation without a
resonance.
The third check is to replace the Breit-Wigner denominator of
Eq. (15) by
\begin {equation}  
D = A - m(s) - iM[\Gamma _1 + \Gamma_2F(s)].
\end {equation}
This removes the resonance by changing $M^2 - s$ to a constant
$A$; it is the term in $s$ which drives the real part of the
amplitude through zero on resonance.
Fig. 12 shows the Argand diagram of the amplitude fitted to
WA102 data.
[The vertical scale is no longer limited to 1 at the peak, because
the imaginary part of the amplitude no longer needs to reach 1
in the absence of a resonance.]

\begin{figure}[htb]
\begin{center}
\vskip -6mm
\epsfig{file=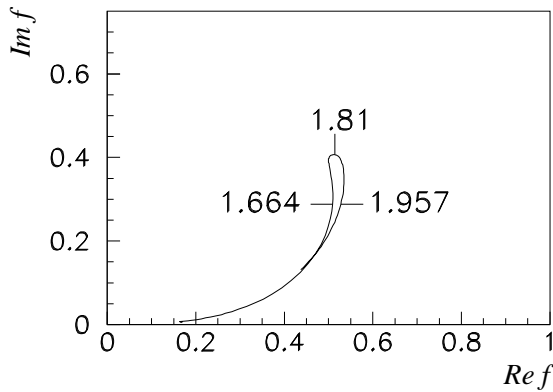,width=8cm}
\vskip -6mm
\caption {The Argand diagram for a fit to WA102 data using Eq.
(26).}
\end{center}
\end{figure}

The imaginary part of the amplitude is forced by the data to peak
at $\sim 1810$ MeV, as before.
The real part is positive everywhere.
It turns out that the data do not force the constant $A$ to go
negative; instead it optimises close to $+0.5$.
Then fits to $\eta 3\pi ^0$ data are worse than for a resonance
by the amounts shown in column 3 of Table 6.
At the lowest three beam momenta, where the $\eta _2(1870)$ is
strongest, results are similar to column 2. At higher momenta,
the changes drop by roughly a factor 2.
The reason for this drop has been traced to the fact that the
phase variation in Fig. 12 adds a degree of freedom in the fit
compared with column 2, where there is no phase variation at all.
However, the significance levels in column 3 still average
9 standard devations per momentum.
If any phenomenologist wishes to develop a more complete
dynamical model, the data are publicly available from the
authors.
Meanwhile, the evidence for resonant behaviour appears to be
strong.

\section {Conclusions}
This work confirms that the $\eta _2(1870)$ has a branching fraction
to $f_2(1270)\eta $ comparable with that to $a_2(1320)\pi$,
in agreement with our earlier analysis.
It is not possible to fit these data with the large branching
fraction found by WA102.
Results of the two experiments agree well for the mass and width
of $\eta _2(1645)$ and the branching fraction of its decays to
$a_0(980)\eta$.
They also agree quite well for the mass and width of $\eta_2(1870)$ from
decays to $f_2(1270)\eta$.
Small branching fractions reported in Table 2, particularly for
decays to $a_0\pi$, have changed significantly from earlier values for
a complex of reasons which are understood.

\begin{figure}[htb]
\begin{center}
\vskip -12mm
\epsfig{file=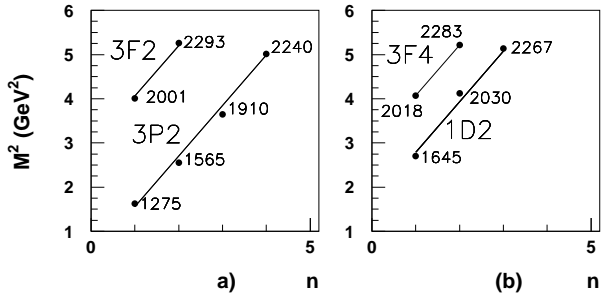,width=10cm}
\vskip -6mm
\caption {Trajectories of resonances for several quantum numbers of
$I=0$, $C=+1$ states; masses are shown in MeV. }
\end{center}
\end{figure}
The WA102 collaboration has found evidence for a weak
decay mode of $\eta_2(1645) \to K\bar K\pi$ \cite {KKpi102} ($7\%$ of
that for decays to $a_2\pi$); however, they find no evidence for
$\eta_2(1870)$ in the same data.
It therefore appears unlikely that the $\eta_2(1870)$ is the $s\bar s$
partner of $\eta_2(1645)$ and $\pi _2(1670)$.
Fig. 13 shows trajectories of $I=0$, $C=+1$ for several sets
of quantum numbers. In (b) $\eta _2(1645)$, $\eta_2(2030)$ and
$\eta_2(2267)$ are consistent with the $^1D_2$ trajectory with the same
slope as the others.

Hybrids with $J^{PC}=2^{-+}$ are predicted around 1900 MeV by Isgur
and Paton \cite {IsgurPat} and Godfrey and Isgur \cite
{Godfrey}.
The interpretation given by many authors, including ourselves
\cite {eta3pi}, is that $\eta_2(1870)$ and $\pi _2(1880)$ make
a hybrid pair, somewhat higher in mass than the $\pi _1(1600)$ of the
PDG, whose average mass is now 1662 MeV.
It would be valuable to search for  $s\bar s g$ partners in $J/\Psi$
decays at BES 3.
Page \cite {Page} predicts that hybrids will decay dominantly to
$^3P$ $\bar qq$ plus a pion, in agreement with the strongest observed
decay modes $a_2\pi$ and $f_2\eta$ of $\eta_2(1870)$.
Predictions for their branching ratio are subject to systematic errors
from (i) the effect of the dispersive attraction from the $f_2\eta$
threshold and (ii) possible mixing of the hybrid with $q\bar q$ 1D and
2S states.
The phase space alone for $a_2\pi$ is $1.184 \pm 0.023$ times that for
$f_2\eta$ after integrating over the line-shape of $\eta_2(1870)$ and
taking account of the form factor $\exp [-(4.5 \pm 1.0)q^2]$.

Li and Wang propose that the $\eta _2(1870)$ is the $n=2$ $q\bar q$
state and $\eta (2030)$ is the $n=3$ state \cite {Wang}.
However, this would required a  trajectory with twice the slope of
other $J^P$.

Afonin \cite {Afonin} has presented an interesting scheme to accomodate
known light mesons.
Its general features are appealing.
However, we question the way $J^{PC} = 2^{-+}$ states are included.
He includes $\eta_2(1645)$, $\eta_2(1870)$ and $\eta_2(2250)$ for
$I=0$ and $\pi_2(1670)$, $\pi_2(2100)$ and $\pi_2(2245)$ for $I=1$.
The large mass difference between $\eta_2(1870)$ and $\pi_2(2100)$ is
unexplained; the mass splitting between $\eta_2(1870)$ and $f_2(1934)$
is reversed for $\pi_2(2100)$ and $a_2(2030)$.

\begin{table}[h]
\begin {center}
\begin{tabular}{cc}
\hline
Momentum (MeV/c) & Change in log likelihood \\\hline
1200 & 408 \\
1350 & 387\\
1525 & 420 \\
1642 & 536 \\
1800 & 484 \\
1940 & 215 \\\hline
\end{tabular}
\caption{Changes in log likelihood when $\eta _2(2030)$ is removed
from the fit to $\eta 3\pi ^0$ data.}
\end {center}
\end{table}
His model conflicts with present data, which require the
presence of $\eta_2(2030)$;
Table 7 shows changes in log likelihood if it is omitted and all
other parameters are re-optimised.
They are on average 1.18 $\times$ values in Table 7 for $\eta_2(2030)
\to [a_2\pi]_{L=2}$, but they are not as large as the sum for all
decays of $\eta_2(2030)$.
This is because of correlations between decay channels.
Nonetheless they are still highly significant.
Furthermore, there is independent evidence for $\eta_2(2030)$ in
$\bar pp \to \eta\pi\pi$ \cite {epp}.
In those data, there is a strong peak with $M = 2040 \pm 40$ MeV,
$\Gamma = 190 \pm 40$ MeV in $[f_2(1270)\eta]_{L=0}$ and a smaller, but
still significant peak in $[f_2(1270)\eta]_{L=2}$;
unfortunately the resonant phase variation cannot be checked in those
data because there is no other strong feature in other singlet states
near this mass.

The $\pi_2(1880)$ is listed by the PDG in four sets of data:
(i) in $\eta\eta\pi$ with $M = 1880 \pm 20$ MeV by Anisovich et al.
\cite {CBAR},
(ii) in $\pi^-p \to \eta\pi^+\pi^-\pi^-p$ with
$M = 2003 \pm 88 \pm 148$ MeV by E852 \cite {Kuhn},
(iii) in $\pi^-p \to \omega\pi^-\pi^0p$ with $M = 1876 \pm 11 \pm 67$
MeV in further E852 data of Lu et al. \cite {Lu}, and
(iv) in $\pi^-p \to \eta\eta\pi^-p$ with $M = 1929 \pm 24 \pm 18$
MeV by Eugenio et al. (E852) \cite {E8523}.
Of these, the second one could be $\pi_2(2005)$.

A further result from the present analysis is that there is evidence
for the presence of $a_2(2255) \to f_2(1270)a_0(980)$
with parameters close to those of Ref. \cite {I1}.
It is the third set of data in which it has been observed, the others
being $\pi \eta$ and $3\pi ^0$.
There is also evidence that the channel $[\eta_2(2030)\pi]_{L=1}$ is
produced via $\pi_2(2245)$; its mass and width are determined better
by present data than by earlier analyses:
$M = 2285 \pm 20 (stat) \pm 25(syst)$ MeV, $\Gamma = 250 \pm 20(stat)
\pm 25(syst)$ MeV.
This is the third channel in which it has been observed.

Further data on $\eta 3\pi ^0$ from a transversely polarised
target would be very valuable.
In such data, there are interferences between singlet and triplet
states.
It is likely that such information would allow a complete spin-parity
analysis of the data, improving further on the present analysis.

\section {Acknowledgements}
We thank the Crystal Barrel Collaboration for the use of the data.

\end  {document}